\documentstyle[12pt]{article}

\def\be{\begin{equation}}
\def\ee{\end{equation}}
\def\ba{\begin{array}{c}}
\def\ea{\end{array}}
\def\p{\partial}
\def\gtrsim{\stackrel{>}{\sim}}
\def\ben{$$}
\def\een{$$}
 
\begin{document}

\titlepage

  \begin{center}{\Large \bf
Fragile ${\cal PT}-$symmetry in a
 solvable model
 }\end{center}

\vspace{5mm}

  \begin{center}

Miloslav Znojil\footnote{ e-mail: znojil@ujf.cas.cz}

 \vspace{3mm}

\'{U}stav jadern\'e fyziky AV \v{C}R, 250 68 \v{R}e\v{z}, Czech
Republic\\

 \vspace{3mm}

\end{center}

\vspace{5mm}

\section*{Abstract}

One of the simplest pseudo-Hermitian models with real spectrum
(viz., square-well on a real interval ${\cal I}$ of coordinates)
is re-examined. A ${\cal PT}-$symmetric complex deformation ${\cal
C}$ of ${\cal I}$ is introduced and shown tractable via an
innovated approach to matching conditions. The result is
surprising: an {\em arbitrarily small} deformation ${\cal I} \to
{\cal C}$ implies a sudden collapse (i.e., the spontaneous ${\cal
PT}-$symmetry breaking) of virtually {\em all} the spectrum (i.e.,
up to its low-energy part).

\vspace{9mm}

\noindent PACS  03.65.Fd

\vspace{9mm}

  \begin{center}
\end{center}

 \newpage

\section{Non-Hermitian Hamiltonians
and their spectra}

Thirty five years ago, Bender and Wu \cite{BW} published an
extremely exciting discovery that certain bound-state problems may
be {much better} understood when one {drops} the ``obligatory"
Hermiticity assumption $H = H^\dagger$ and admits that a coupling
constant $g>0$ in Schr\"{o}dinger equation
 \be
 H(g)\,|\psi_n(g)\rangle =
 E_n(g)\,|\psi_n(g)\rangle, \ \ \
  \ \ \ \ \ \ \ \ n = 0, 1, \ldots
 \ee
is analytically continued to a complex value $g \in l\!\!\! C$. In
this perspective Bender and Wu worked, for definiteness, with the
quartic anharmonic-oscillator Hamiltonians
 \be
 H(g) =H^{(4)}(g) = \hat{p}^2
 + f^2\hat{x}^2 + g\,\hat{x}^4,
 \label{quartic}
 \ee
and demonstrated that the separate (though, in general, complex)
spectra $\{ E_n(g)\}$ may {\em all} be interpreted as the sets of
an intersection of {\em all} the Riemann sheets of a {\em single}
analytic function $I\!\!E^{(4)}(g)$ with a corresponding ``line"
of a constant $g$. Many years later, similar observations were
made and verified for the cubic model
 \be
 H(g) =H^{(3)}(g) = \hat{p}^2
 + f^2 \hat{x}^2 + i\,g\,\hat{x}^3\,
 \label{cubic}
 \ee
(see the review \cite{Alvarez} for more details) etc. In all of
these models, the costs of the generalization $H \neq H^\dagger$
proved much lower than expected. For all the nonzero couplings
$g\neq 0$, all of their complex ``exceptional points" (EP)
\cite{Heiss} proved well separated in the complex plane of $g$ for
both $m=4$ and $m=3$.

A new important development of the subject emerged cca six years
ago when Bender and Boettcher published their letter \cite{BB}.
Having extended their attention to the whole class of the
power-law models
 \be
 H(g) =H^{(2+\delta)}(g) = \hat{p}^2
 +  f^2 \hat{x}^2 + g\,\hat{x}^2\,\left (i\,\hat{x}
  \right )^\delta, \ \ \ \ \ \ \ \ \delta > 0,
  \ \ \ \ \ \ \ \ g > 0\,
 \label{gecubic}
 \ee
[reproducing the above special cases (\ref{quartic}) and
(\ref{cubic}) at $\delta = 2$ and $\delta = 1$, respectively],
they summarized several existing perturbative and numerical
experiments (for illustration one could cite, e.g.,
\cite{Caliceti,DB,BG}), complemented them by numerous new WKB
arguments and conjectured that after one introduces a suitable
Hilbert space, {\em all} the Hamiltonians (\ref{gecubic}) (with,
for simplicity, $f=0$) will possess the {\em purely real} (and
discrete and, from below, bounded, i.e., ``observable-like")
spectra, in spite of their {\em manifestly} non-Hermitian origin.

For the pioneering conjecture of this type, several rigorous
proofs \{cf., e.g., the Fourier-transformation results \cite{BG}
for eq. (\ref{quartic}) at $g<0$, or the manifest reality of
perturbative energies after re-summation for eq. (\ref{cubic}) at
the real $g$ \cite{Caliceti}\} were already available and many
other had only to come \{cf.,  for illustration, proofs in the
difficult case of eq. (\ref{cubic}) in \cite{DDT}\}. Nevertheless,
the core of the message delivered by Bender and Boettcher lied in
the emphasis attributed to the parity-plus-complex-conjugation
symmetry (conveniently called ${\cal PT}-$symmetry) of their
sample non-Hermitian Hamiltonians $H^{(2+\delta)}(g)$ with real
spectra. This inspired an extensive subsequent study of the
structure of the relationship between the reality of the spectrum
and the ${\cal PT}-$symmetry of the underlying non-Hermitian
Hamiltonian \cite{ostatni}.

We intend to contribute to the latter effort by the description of
an exactly solvable example which exhibits a rather
counterintuitive enhanced sensitivity to a very small change of
its coordinate domain. We shall start from an overall review of
the state of the art in section \ref{sectiondve} where we
emphasize the theoretical importance as well as some practical
weaknesses of the pseudo-Hermitian constraint imposed upon the
non-Hermiticity of the Hamiltonians.

In the next section \ref{sectiontri} we return to the study of
quantitative characteristics of the specific,
differential-equation models where the current and robust property
of Hermiticity $H=H^\dagger$ is being replaced by the ${\cal
PT}-$symmetry which may be fragile \cite{BBjmp,Lev}. We restrict
our attention to the most elementary non-Hermitian square-well
model (of ref. \cite{SQW}, with real spectrum) and extend its
scope slightly by the replacement of its usual domain (viz., a
finite interval ${\cal I}$) by a broken line (or by any other
smoothly deformed curve ${\cal C}$ - cf. Figure 1) in the complex
plane of coordinates.

Our main mathematical results are presented in section \ref{IV}
where our new method of solving the matching conditions is shown
applicable to an explicit qualitative description of the structure
of the bound states in the broken-path regime. We explain in
detail how our ``moving-lattice" method decisively facilitates the
global analysis of the matching conditions.

Our key physical message is finally formulated in section~\ref{V}
emphasizing that our innovative geometric interpretation of the
matching conditions offers the rigorous proof that in our example,
an {arbitrarily} small imaginary shift $0 \to i\,\omega$ of the
matching point causes a non-perturbative, {\em sudden}
complexification of {\em all} the high-energy part of the
spectrum. In the currently accepted terminology this means that
the vast majority of the wavefunctions encounters a spontaneous
breakdown of {\em their} ${\cal PT}-$symmetry. We shall conclude
in section \ref{VI} that this symmetry is {\em manifestly} fragile
in our particular model.

\section{ ${\cal PT}-$symmetry
\label{sectiondve}}

\subsection{Variational picture \label{subsectiondve}}

In the early stages of study of ${\cal PT}-$symmetric quantum
mechanics people tried to understand the complexified
Schr\"{o}dinger Hamiltonians of the type (\ref{gecubic}) as models
on the real line, with a pair of the real and imaginary potentials
of a definite behaviour with respect to the parity ${\cal P}$.
Along these lines one arrives at an introduction of the two
harmonic-oscillator-type bases $\{\,|n^{(\pm)}\rangle \,\}$ (with
definite, fixed parities $^{(\pm)}$) and transforms the ${\cal
PT}-$symmetric differential Schr\"{o}dinger equations
$H\,|\psi\rangle=E\,|\psi\rangle$ with a pre-selected
normalization of $|\psi\rangle =\sum_n \left ( |n^{(+)}\rangle
\psi^{(+)}_n + i\,|n^{(-)}\rangle \psi^{(-)}_n \right ) $ into the
variational-like real and partitioned matrix problems containing
arrays $\vec{\psi}^{(\pm)}$ of the real wavefunction components
${\psi}^{(\pm)}_n$,
 \be
 \left (
 \begin{array}{cc}
 A&-C\\
 C^T&D
 \ea
 \right )
 \left (
 \begin{array}{c}
 \vec{\psi}^{(+)}\\
 \vec{\psi}^{(-)}
 \ea
 \right )=
 E\,\left (
 \begin{array}{c}
 \vec{\psi}^{(+)}\\
 \vec{\psi}^{(-)}
 \ea
 \right )\,.
 \label{ioch}
 \ee
The infinite-dimensional submatrices $A=A^T$ and $D=D^T$ are real
and symmetric but the spectrum itself need not be real at all
\cite{web}. These considerations inspired Mostafazadeh who
conjectured, in a series of papers \cite{AM}, that the Bender's
and Boettcher's ${\cal PT}-$symmetric quantum mechanics should be
classified as a mere special case of the more universal
pseudo-Hermitian quantum mechanics, the origins and foundations of
which might be traced back to Dirac et al \cite{Dirac}. In such an
overall setting, he proposed to weaken the ${\cal PT}$ symmetry of
the Hamiltonians to their mere pseudo-Hermiticity
 \be
 H^\dagger = \eta\,H\,\eta^{-1}, \ \ \ \ \ \ \eta=\eta^\dagger
 \label{constraint}
 \ee
where we may set, in our particular example (\ref{ioch}),
 \be
 \eta=\eta_{\cal P}=
 \left (
 \begin{array}{cc}
 I&0\\
 0&-I
 \ea
 \right )=\eta_{\cal P}^{-1}\,.
 \label{paritne}
 \ee
In the light of the well known \cite{Geyer} huge ambiguity of the
assignment of the ``metric" $\eta$ to any given pseudo-Hermitian
Hamiltonian $H \neq H^\dagger$, Mostafazadeh also proposed that
the ``natural" choices with the indeterminate parity-like metric
operators [like $\eta_{\cal P}$ in eq. (\ref{paritne})] should be
{\em all} replaced by {\em any} (i.e., very often, non-diagonal
and strongly Hamiltonian-dependent) positive definite alternative
$\eta_{+}>0$. In parallel to Mostafazadeh, similar conclusions
have been reached in refs. \cite{ptho} and \cite{BBJ} defining, in
the present language, the particular positive definite metrics
$\eta_+={\cal PQ}$ and $\eta_+={\cal CP}$ using the additional
symmetry generators ${\cal Q}$ and ${\cal C}$ of quasi-parity and
charge, respectively.

The latter procedure enables us to call al the similar
non-Hermitian Hamiltonians $H$  ``quasi-Hermitian" because, in
accord with the review \cite{Geyer}, the positivity of
$\eta_{+}>0$ suppresses many interpretation difficulties and
leaves the quasi-norm $||\varphi||=\sqrt{\langle \varphi |
\eta_{+} | \varphi \rangle}$ real and non-degenerate. This makes
the corresponding Hamiltonians compatible with their standard
quantum-mechanical probabilistic tractability~\cite{AMC}.

\subsection{Square-well illustration
 \label{oldccsquare}}

One of the main sources of inspiration for the selection of
potentials in  Schr\"{o}dinger equations (say, in the Coulombic
form) is the principle of correspondence which allows us to extend
and transfer to quantum mechanics the experimental experience
gained during centuries in the common, macroscopic world. A
counterintuitive character of many quantum phenomena allows us to
search for some new and unusual Schr\"{o}dinger equations, e.g.,
by a complexification of their axes of coordinates $I\!\!R \to
{\cal C}$ \cite{BB}. Alternatively, we may obtain manifestly
${\cal PT}-$symmetric equations
 \be
 \left [ -\frac{d^2}{dx^2} + V(x) + i\,W(x)
 \right ]\,
 \psi(x)=E\,\psi(x)
 \label{basic}
 \ee
by staying on the real line and ``deforming" the shapes of
$V(x)=+V(-x)$ and $W(x)=-W(-x)$. Samples of both these approaches
may be found, e.g., in ref. \cite{BBjmp} (considering mainly the
asymptotically power-law potentials) or \cite{Fern} (paying
attention to some exponentially confining forces) or \cite{SQW}
(where an even steeper, infinitely deep square well has been
complexified) or \cite{Albeve} (where the mathematical properties
have been discussed for the next-step models with delta-functions
mimicking the ``infinitely thin" square wells). The square-well
model represents a reasonable phenomenological compromise
exhibiting, as a bonus, an important merit of exact solvability.
We shall pick up it in what follows, interpreting the infinitely
deep real part of the potential
 \be
  V(x) = \left \{  \begin{array}{l} + \infty\\ \ \  0 \\
 + \infty \ea \right . \ \ \ \ \ \ \ {\rm for\ \  \ } \ \ \
 \left \{
  \begin{array}{l}
 \ x\
 > 1 \\
  -1 < x
 < 1 \\
 \ x\
 < -1
   \ea
   \right .
 \label{SQW}
  \ee
as a requirement that all the wavefunctions vanish at $x = \pm 1$.
We break the Hermiticity of our Hamiltonian by adding the
imaginary and ${\cal PT}-$symmetric finite interaction term with
coupling $Z>0$,
 \be
  W(x) = \left \{  \begin{array}{l} + iZ,\\ -
 iZ \ea \right . \ \ \ \ \ \ \ {\rm for\ \  \ } \ \ \
 \left \{
  \begin{array}{l}
 {\rm Re}\ x\
 < 0 \\
 {\rm Re}\ x\
 > 0
   \ea
   \right ..
 \label{SQWe}
  \ee
Once we assume that the interval of the coordinates $x$ remains
purely real, the spectrum of energies $E_n=E_n(Z)$ proves discrete
and real at all $Z < Z_{critical} \approx 4.48$ (cf. ref.
\cite{Geza}). It smoothly converges towards the well known
square-well energy levels $E_n(0)=(n+1)^2\pi^2/4$ in the Hermitian
limit $Z \to 0$.

In an extension of the above square-well model we shall now assume
that the interval of the coordinates $x$ is deformed to complex
plane. The corresponding generalized, ${\cal PT}-$symmetric (i.e.,
left-right--symmetric) curve ${\cal C}$ is sampled in Figure~1.

\section{Exact solvability of the new model \label{sectiontri}}

Once we define our potential (\ref{SQWe}) in the whole complex
plane of $x \in l\!\!\!C$, solutions $\psi(x)$ will be analytic in
both its half-planes. The only distinctive feature of our present
generalization $(-1,1)= {\cal I} \longrightarrow {\cal C}$ lies in
the requirement of the matching of the left and right branches
$\psi_\mp (x)$ of our full wavefunction $\psi(x)$ at a point
$x_0=i\,\omega$ on the imaginary axis. This enables us to
postulate the matching rules
 \be
 \psi_-(i\,\omega)=\psi_+(i\,\omega) = 1, \ \ \ \ \ \
 \p_x\psi_-(i\,\omega)=\p_x\psi_+(i\,\omega) = i\,A
 \label{conditor}
  \ee
in terms of an auxiliary real parameter $A \in (-\infty,\infty)$.

\subsection{Re-parametrization of the matching conditions}

As long as our potentials $V$ and $W$ are constant almost
everywhere, the general solution of our differential
Schr\"{o}dinger equation (\ref{basic}) may be put equal to a sum
of the hyperbolic sine and cosine. The left and right solutions
$\psi_\mp (x)$ are different, having to vanish at the different
boundary points $x \to \mp 1$,
 \be
  \psi_-(x) = R_-\,\sinh \kappa^*(1+x),\ \ \ \ \ \ \
   \psi_+(x) = R_+\,\sinh \kappa(1-x)\,.
 \label{ansatz}
 \ee
With $\kappa= s - i\,t$, the values of the two free real
parameters $s $ and $t $ will be determined by the differentiation
in eq. (\ref{basic}),
 \be
  E = t^2-s^2, \ \ \ \ \ \ \ Z = 2\,s\,t.
  \label{tworel}
 \ee
As long as a change of the sign of $\kappa$ would influence just
the (arbitrary) sign of the overall normalization coefficients
$R_\pm$, we conveniently restrict our attention to the quadrant of
$s>0$ and $t>0$ (note that we fixed the sign of $Z> 0$ in
advance). The insertion of the right and left solutions
(\ref{ansatz}) in the matching conditions (\ref{conditor}) gives
the following complex (and transcendental) algebraic equations,
  \ben
   L\,\sinh
 \kappa^*(1+i\,\omega)= R\,\sinh \kappa(1-i\,\omega)=1, \een \ben
 \ \ \ \
 \kappa^*
L\,\cosh \kappa^*(1+i\,\omega)= -\kappa\, R\,\cosh
\kappa(1-i\,\omega)=i\,A.
 \een
Their solution is our main task. In the first step, we can get rid
of the redundant constants by taking the ratios,
 \be
 \kappa^*\,
{\rm cotanh}\, \kappa^*(1+i\,\omega)= -\kappa\
 {\rm cotanh}\,
 \kappa\,(1-i\,\omega)=i\,A.
 \label{tojeon}
  \ee
The former equal sign is trivial while the latter one represents a
complex equation which defines the real parameter $A$ {\em and}
inter-relates the two real and positive parameters $s$ and $t$ in
addition. As long as we have $Z=2st$, this should determine all
their admissible values.


Changing our notation and putting $\omega= \tan \varphi$ with
$\varphi \in -\pi/2,\pi/2)$, let us now introduce two auxiliary
linear functions $S=S(s,t)$ and $T=T(s,t)$ defined by the
elementary two-dimensional rotation
 \be
 \left (
 \ba
 S\\T
 \ea
 \right )
 =
 \left (
 \begin{array}{cc}
 \cos \varphi & -\sin \varphi\\
 \sin \varphi&\cos \varphi
 \ea
 \right )
 \,
 \left (
 \ba
 s\\t
 \ea
 \right )\,
 \label{op}
 \ee
where the angle of the rotation measures also the upward shift of
the matching point in the complex plane of our complex
coordinates. In this notation we may re-write our matching
constraint (\ref{tojeon}) in the form
 \be
 A\,\sinh
 \left (
 \frac{S}{\cos
 \varphi} - i\, \frac{T}{\cos
 \varphi}
 \right )=(t+i\,s)\,\cosh
 \left (
 \frac{S}{\cos
 \varphi} - i\, \frac{T}{\cos
 \varphi}
 \right )
 \label{matchch}
 \ee
which admits a facilitated separation of its real and imaginary
part. The value of $A$ drops out of their ratio which may be
further re-arranged to represent our matching condition in the
most compact real form
 \be
 s\,\sinh \left (
 \frac{2S}{\cos \varphi}
 \right )=-t\,\sin \left (
 \frac{2T}{\cos \varphi}
 \right )
 . \label{implicit}
 \ee
In the limit $\varphi \to 0$ the latter equation coincides with
the $\omega=0$ prescription of ref. \cite{SQW}. At the generalized
$\omega \neq 0$ the replacement of $s$ and $t$ by $S$ and $T$ via
eq. (\ref{op}) converts our new and more complicated matching
formula (\ref{implicit}) into its final form
 \be
 \tau = \sigma \,\frac{\omega + \varrho(\tau)\,\sinh \sigma}
 {1-\varrho(\tau)\,\omega\,\sinh \sigma}, \ \ \ \ \ \ \ \
 \omega = \tan \varphi
 \label{19}
 \ee
where we abbreviated $\sigma = 2S/\cos \varphi$, $\tau = 2T/\cos
\varphi$ and $\varrho = \varrho(\tau) =-1/\sin \tau$. This
equation is an implicit definition of a certain set of curves
$\tau=\Theta(\sigma)$ in the $\sigma-\tau$ plane. In principle,
the knowledge of these curves would enable us to find all their
intersections $(\sigma_k,\tau_k)$, $ k = 0, 1, \ldots$ with our
original constraint $t=Z/(2s)$.

\subsection{The lattice-moving method of solving eq. (\ref{19})
 \label{lattshif} }

Our present key idea is that the function $\varrho = \varrho
(\tau)$ is periodic, i.e., it {remains constant} on a discrete
lattice ${\cal L}$ of its argument $\tau$. In this spirit we shall
split the real axis of $\tau$ into intervals of the length $2\pi$
numbered by an integer $k$. Then we introduce the second variable
$p = \pm 1$ marking the right and the left half of each of these
intervals, respectively. This guarantees that at a fixed $p$ the
sign of the sine function remains the same and equal to $-p$.
Finally, due to the symmetry of each of the sine-shaped curves we
split the half-intervals in the quarter-intervals marked by
another index $q=\pm 1$,
 \be
 \tau =(2k+1)\,\pi
 + p\,\frac{\pi}{2} + q\,\frac{\pi}{2}\,\xi
 \equiv
 \tau_{(k,q)}(p,\xi)\,,\ \ \ \ \ \ \ \xi =\xi(tau) \in (0,1).
 \label{diskr}
 \ee
As a consequence, our parameters $\varrho(\tau)$ become
represented by the functions which are independent of $k$ and $q$,
 \be
   \Omega(p,\xi) = -\frac{1}{\sin \tau_{(k,q)}(p,\xi)}
  = +\frac{p}{\cos (\pi \xi/2)}\,.
  \label{defom}
  \ee
This means that the parameters $\varrho$ remain constant over all
the lattices ${\cal L}_{(p_0,\xi_0)}$ of points $
\tau_{(k,q)}(p_0,\xi_0)$ where the sign $p_0$ and the parameter
$\xi_0$ are temporarily fixed.

\subsubsection{Verification: Straight-path solution
re-visited \label{4.1.1}}

At $\omega=0$ and  ${\cal C} = {\cal I}$, the use of the limiting,
simplified version
 \be
 \tau = \varrho(\tau)\,\sigma \,{ \sinh \sigma}
 \,, \ \ \ \ \ \ \ \omega=0
 \label{quickly}
 \ee
of our matching condition (\ref{19}) leads to an enormous
simplification of the construction performed in ref. \cite{SQW}.
There, severe difficulties originated from a strong and pronounced
$\tau-$dependence of the factor $\varrho = \varrho (\tau)$ which
is a very quickly changing function of its argument $\tau$. In our
present setting, the discretization (\ref{diskr}) enables us to
fix the value of $\varrho = \Omega$ by reducing our attention from
all the values of $\tau$ to their lattices ${\cal L}_{(p,\xi)}$.
Treating them separately, one at a time, we only have to keep in
mind the overall range of our real constants $ \Omega(p,\xi)= p\,
| \Omega(p,\xi)|= p\, \Omega(+1,\xi)  \notin (-1,1)$. This enables
us to re-parametrize the matching condition (\ref{quickly}),
 \be
 \tau = \tau_{(k,q)}(p,\xi)=\Omega(p,\xi)\,\sigma \,{ \sinh \sigma}
 \,, \ \ \ \ \ p, \xi = fixed\,.
 \label{quicklynu}
 \ee
In the new language, the graph of the function $\Omega\,\sigma \,{
\sinh \sigma}$ is a parabolic curve which is oriented up or down
at the respective $p=+1$ and $p=-1$. As long as we are interested
in the positive $\tau>0$, we may discard $p=-1$ and fix $\tau =
\tau_{(k,q)}(+1,\xi)>0$ and $\varrho(\tau)
=\Omega(p,\xi)=\Omega(+1,\xi)\geq 1$. The curves $\tau
=\Omega(+1,\xi)\,\sigma \,{ \sinh \sigma} \equiv \Theta_{\xi}
(\sigma)$ then shrink in proportion to the growth of $\xi$,
proceeding from their broadest $\xi=0$ version (where $|\Omega| =
1$) via the narrowing parabolic curves until the degenerate single
and upwards-oriented half-line in the limit $\xi \to 1$, i.e.,
$\Omega \to \infty$. This is illustrated in Figure 2 where the
unlimited shrinking of the curves is sampled at $\xi=0, 0.5, 0.9 $
and $\xi = 0.99$.

When we zoom out a  stripe of $\tau=\tau_{(k,q)}(p,\xi)$ at a
fixed $k=30$ in Figure 2, we get Figure 3. In the new Figure the
variations of $\tau$ are determined solely by the changes of $\xi$
and $q$ which are sampled by a few horizontal lines. As long as
the right-hand-side function $\Theta_{\xi} (\sigma)$ depends on
both $\sigma$ and $\xi$, equation (\ref{quicklynu}) will be
satisfied {\em only} at the points of intersection of each
particular $\xi-$marked horizontal line with another particular,
$\xi-$assigned parabolic curve. In this manner the points of
intersection $(\sigma_{\xi_m},\tau_{\xi_m})$ in Figure 3 sample
the graphical solution of the matching condition
(\ref{quicklynu}). As long as we choose a fairly large stripe
number $k = 30$, the parabolas of Figure 2 are represented by the
almost straight and almost vertical lines in Figure 3. This makes
the identification of all the intersections particularly easy. We
see that the points of intersection form the horizontally prolate
ovals, each of which being confined within its $k=k_0$ and $p=+1$
stripe, and not exceeding the interior of the ``maximal",
$|\Omega|=1$ parabola. It is obvious that the horizontal lines (=
lattices of $\tau$) as well as the more or less vertical parabolas
move smoothly with the growth of $\xi$. The resulting picture
reproduces precisely our old graphical proof \cite{SQW} of the
existence of solutions at $\omega=0$. Our present new
discretization method appears to offer a feasible extension of
this proof and analysis to $\omega \neq 0$.

\section{Bound states at $\omega \neq
0$ in graphical representation
 \label{IV}}

Once we wish to determine the spectrum of the square-well energies
$E_n=t_n^2-s_n^2$ at any $\omega \neq 0$, we have to find all the
real values of $s=s_n$ and $t=t_n$ which satisfy {\em both} the
constant-coupling constraint (\ref{tworel}) {\em and} the matching
condition (\ref{implicit}). In the first step, let us re-express
the former elementary hyperbolic-curve correlation $Z = 2st$ in
the new variables $\sigma$ and $\tau$.

\subsection{The $Z-$dependent hyperbolic-curve constraint }

Rotation (\ref{op}) implies that under the assumption $\omega>0$
we have $\tau > 0$ while the sign of $\sigma$ may be both positive
and negative. Alternatively, the choice of $\omega < 0$ would
imply that we must keep $\sigma > 0$ while the sign of $\tau$ is
allowed to vary. This means that one of the two hyperbolas defined
by the rule $Z = 2st$ may be discarded immediately. Of course, in
our innovated notation we must describe these hyperbolas by the
slightly less transparent rotated quadratic equation
 \be
 \tau^2 + 2\tau\sigma\,\frac{\cos 2\varphi}{\sin 2\varphi}
 - \sigma^2- \frac{4Z}{ \sin
 2\varphi\,\cos^2\varphi}=0.
 \label{ksindl}
 \ee
At $\omega=\tan \varphi >0$ it is easy to select the correct
branch defined by the formula
 \be
  \tau=
 \Xi(\sigma)=
  \frac{1}{2}\,
 \left (
 \omega-\frac{1}{\omega}
 \right )\,\sigma  +\frac{1}{2}\,\sqrt{
 \left (\omega+\frac{1}{\omega} \right )^2 \sigma^2+
   4X^2}\,, \ \ \ \ \ \  X^2= \frac{4Z}{ \sin
 2\varphi\,\cos^2\varphi}\,.
   \label{ksindlta}
  \ee
In parallel, at $\omega=\tan \varphi= -\tilde{\omega} < 0$ we must
use the {\em different}  formula
 \be
  \sigma=
 \Upsilon(\tau)=
  \frac{1}{2}\,
 \left (
 \frac{1}{\omega}-\omega
 \right )\,\tau  +\frac{1}{2}\,\sqrt{
 \left (\omega+\frac{1}{\omega} \right )^2 \tau^2+
   4Y^2}\,, \ \ \ \ \ \ Y^2= \frac{-4Z}{ \sin
 2\varphi\,\cos^2\varphi}\,.
   \label{ksindlsb}
  \ee
In the other words, we must treat the up and down shifts
$i\,\omega$ of the matching point separately, reflecting the fact
that we already broke the symmetry between the half-planes of
coordinates $x \in l\!\!C$ by having chosen the positive coupling
$Z>0$ in advance.

\subsection{The second, matching constraint}

The application of the lattice-shifting technique of section
\ref{lattshif} may be extended to both the positive and negative
$\omega$. The variable $\tau=\tau_{(k,q)}(p,\xi) $ remains
represented by the same function of the interval selector $k$, of
the two sign-variables $q = \pm 1$ and $p = \pm 1$ and of the
continuous $\xi$ varying in the compact interval $(0,1)$.

\subsubsection{Moving lattices \label{themethod}}

Let us now select $\omega>0$ and keep the two auxiliary variables
$p$ and $\xi$ fixed. This restricts the range of our variable
$\tau$ to the lattice ${\cal L}={\cal L}(p,\xi)$ where the
function $\Omega = \Omega(p,\xi) ={p}/{\cos (\pi \xi/2)}$ remains
constant. This leads to a decisive simplification of our matching
condition (\ref{19}),
 \be
 \tau=\Theta_{(p,\xi)}(\sigma)
  = \sigma \,\frac{\omega + \Omega\,\sinh \sigma}
 {1-\Omega\,\omega\,\sinh \sigma} \,, \ \ \ \ \ p, \xi = fixed\,.
 \label{Om19}
 \ee
A typical graph of the function $\Theta_{(p,\xi)}(\sigma)$ at both
$p=\pm 1$ and at the minimal $\xi =0$ and/or $|\Omega|=1$ is
displayed in Figure 4.  With respect to the growth of the
lattice-characterizing parameter $\xi$ from $0$ to $1$ it is
trivial to see from eq. (\ref{Om19}) that

\begin{itemize}

\item at  $p=+1$ and $\sigma > 0$, the right branch of the
well-shaped curve $\Theta_{(+1,\xi)}(\sigma)>0$ is bounded by its
perpendicular asymptote at $\sigma_\infty(\xi)= {\rm arcsinh}
(1/[\omega\Omega(+1,\xi)])$. With the growth of $\xi$ and $\Omega$
it inadvertently  moves to the left and in the limit of $\xi \to
1$ it coincides with the vertical half-axis $V_{right}$
($\sigma_\infty(1)=0$);

\item
in parallel, the left branch of the same well moves upwards and
coincides with its diagonal asymptote $D_{left}$ in the same
limit, $\lim_{\xi \to 1}\Theta_{(+1,\xi)}(\sigma) =
-\sigma/\omega$;

\item at  $p=-1$ and $\sigma < -\sigma_\infty(\xi) < 0$, there
exists another hyperbolic well sampled in Figure 4, with
definition $\tau=\Theta_{(-1,\xi)}(\sigma)$ and asymptotes
$D_{left}$ and ($\xi-$dependent) $A_{right}$. With the growth of
$\xi$ this well moves downwards and to the right and coincides
with the wedge formed by $D_{left}$ and $V_{right}$ at $\xi = 1$.

\end{itemize}

 \noindent
This geometric picture has several consequences. The most
important one is that at the minimal $\xi=0$ the curves of
Figure~4 contain the initial points of all the ovals of the
solutions in the manner indicated by the $\xi=0$ point in Figure 3
above. With the growth of $\xi$ the similar oval-shaped curves are
then being formed at any $\omega$.

\subsubsection{Four families of half-ovals }

In a continuing description of the structure of solutions of eq.
(\ref{Om19}) we must distinguish between the positive and negative
$\sigma$. For $\sigma \geq 0$, the analysis is simpler since the
ovals (or rather half-ovals) as sampled in Figure 3 at $\omega=0$
can solely exist in the stripes with $p=+1$. With the growth of
$\xi$ they open their two $q=\pm 1$ branches to the left until
they attain their maximal width and reach the boundaries of their
stripes on the vertical axis $V_{right}$ in the limit $\xi \to 1$.

At $\sigma < 0$ we have to parallel the above half-ovals by their
$p=+1$ partners which start to open to the right at the leftmost
curve with $\xi = 0$. They end their growth at $\xi=1$ while
touching the boundaries of their $p=+1$ stripes on the diagonal
$D_{left}$.

In contrast to our above $\omega=0$ exercise in section
\ref{4.1.1}, the choice of $\sigma < 0$ admits the existence of
another family of the half-ovals within the $D_{left}-V_{right}$
wedge. Of course, they can only exist within the stripes where
$p=-1$ and in the domain of the sufficiently large $\tau \gtrsim
\tau_0$ (i.e., at $k \geq k_{minimal}$) where they can originate
on the curve $\tau=\Theta_{(-1,0)}(\sigma)\geq \tau_0$. In this
domain they form the two subfamilies again, depending on whether
they originated on the left or right branch of the $\xi=0$ curve.

\subsubsection{Two patterns of gluing the half-ovals}

With the growth of $\xi$, the left half-ovals within the wedge
$D_{left}-V_{right}$ open to the left, ending their growth at
$\xi=1$ in the intersections of the boundaries of their $p=-1$
stripes with the left diagonal straight line $D_{left}$. At these
points these half-ovals meet their $p=+1$ partners so that in
contrast to the $\omega=0$ pattern (with a series of the separated
and closed ovals -- cf. their picture in ref. \cite{SQW}), the
resulting locus of the solutions forms a wavy, sine-like-shaped
line which oscillates to the left and right and moves up to the
left along the diagonal $D_{left}$. As long as this curve remains
confined between its two envelopes $\Theta_{(\pm 1,0)}(\sigma)$,
the asymptotic decrease of the amplitude of this wobbling is
exponential. A schematic example of such a wavy curve appears in
Figure~5.

In the same range of the sufficiently large $\tau$, the second,
similar wavy pattern is formed along the axis $V_{right}$. In
exactly the same manner it results from the gluing of the right
$p=-1$ half-ovals which open to the right and reach the line
$V_{right}$ where they find a continuation in the above-mentioned
$p=+1$ half-ovals at $\sigma > 0$. In contradistinction to the
previous case, the amplitude of the wobbling is asymptotically
constant. Still, this fact alone is sufficient to exclude this
branch from further consideration because the $Z=2st$ constraint
is asymptotically a hyperbola with asymptotes at the angles
$\varphi = \arctan \omega $ and $\varphi' = \arctan \omega -\pi/2$
with respect to the axis $V_{right}$.

In all the remaining domain of the not too large values of $\tau$,
just smooth perturbations ocur of the $\omega=0$ pattern of
disconnected ovals. At $\omega > 0$ the height of the ovals
exceeds the height of a single stripe. This is consistent with the
fact that an inner part of the ovals lies within the
$D_{left}-V_{right}$ wedge and must belong, therefore, to a $p=-1$
stripe. This also does not contradict to the steady decrease of
the minimum of the graph of the curve $\Theta_{(-1,\xi)}(\sigma)$
since with the growth of $\xi$ the solutions of eq. (\ref{Om19})
start to exist in the lower and lower $p=-1$ stripes within the
wedge. In this way, the resulting loci of solutions of eq.
(\ref{Om19}) are allowed to form the separate ovals in a fully
consistent manner, indeed.

\section{Energies \label{V}}

The ultimate goal of our considerations is achieved. We clarified
that an optimal strategy of the determination of all the
parameters $s=s_n$ and $t=t_n$ in the bound-state formula $E_n =
t_n^2-s_n^2$, $n = 0, 1, \ldots$ is based on a suitable change of
variables $(s, t) \to (\sigma,\tau)$ which merely re-scales and
rotates the original hyperbolic constraint $Z = 2st$ and
re-expresses all the real deformed-path square-well energies by
the ``rotated" formula
 \be
 E=E_n(Z,\varphi)=\frac{1}{4}\,
 \left [
 \left (
 \tau_n^2-\sigma_n^2
 \right )\,\cos 2 \varphi - 2\,\sigma_n\tau_n \,\sin 2\varphi
 \right ]\,\cos^2\varphi\,.
 \label{specg}
 \ee
This leads to a vital simplification of the matching of
wavefunctions. In the real $\sigma - \tau$ plane, the construction
of all the physical bound states (if any) is reduced to an
identification of all the admissible parameters
$(\sigma_n,\tau_n)$ with all the intersections of a certain pair
of curves. One of them is the elementary $Z-$dependent hyperbola
(the smooth curve in Figure 5). A sufficiently transparent
graphical representation of the shape of the second one is more
difficult and required in fact the greater portion of our previous
text. This curve is sampled by its quickly oscillating asymptotic
part in Figure 5.

\subsection{A comment on asymmetry between $\omega > 0$ and $\omega < 0$  }

We mainly paid attention to the positive values of the shift
$\omega>0$ pertaining to the generic form of the family of the
hyperbolae given by eq. (\ref{ksindlta}). They are sampled by the
smoother curve in Figure 5. The Figure also illustrates a generic
pattern of the intersection of these hyperbolae with half-oval
families confined within areas specified by their envelope curves
exemplified in Figure~4.

We did not notice in section \ref{themethod} that after reflection
of Figure~4 with respect to the origin of coordinates $\sigma$ and
$\tau$,  its $p=-1$ and $p=+1$ envelope curves are mapped upon
each other. This simplifies marginally the construction and
follows from the invariance of eq. (\ref{Om19}) {\em on the
lattices} since the simultaneous replacements $\tau \to -\tau$ and
$\sigma \to -\sigma$ are equivalent to $\Omega \to -\Omega$ while
the latter change of sign merely means that we have to transform
$p \to -p$.

We did not deduce, {\it ibidem}, that another simultaneous
sign-change of $\sigma \to -\sigma$ {\em and} $\omega \to -\omega$
preserves the form of the original, lattice-independent matching
condition (\ref{19}). This is more important because at the
negative $\omega = -\tilde{\omega} < 0$ we would be forced to
replace the most complicated pattern of Figure 4 (where we always
employ the positive $\tilde{\omega}=|\omega|$) by its
left-right-reflected copy complemented by the corresponding
correct hyperbolic branch of curve $Z=2st$ in its alternative form
(\ref{ksindlsb}). After the left-right mirroring transform it
enables us to simplify the situation by returning to the original
Figure 4 complemented by the trivially modified {reflected}
hyperbola
 \be
  \sigma=\Sigma
 (\tau)=
  \frac{1}{2}\,
 \left (
 \frac{1}{\tilde{\omega}}-\tilde{\omega}
 \right )\,\tau  -\frac{1}{2}\,\sqrt{
 \left (\tilde{\omega}+\frac{1}{\tilde{\omega}} \right )^2 \tau^2+
   4Y^2}\,, \ \ \ \ \ \ Y^2= \frac{+4Z}{ \sin
 2\tilde{\varphi}\,\cos^2\tilde{\varphi}}\,.
   \label{ksindlbs}
  \ee
Hence, all what we have derived at the positive $\omega$ may {\em
immediately} be transferred to the case where $\omega$ is
negative, {\em without} changing the half-oval curves and with the
mere {addition} of the second branch (\ref{ksindlbs}) of the
hyperbola. In Figure 5 this would just mean a replacement of the
upper hyperbola by its minus-sign partner. Of course, such an
extension of the whole picture is essentially trivial and it need
not be discussed separately at all.

\subsection{The breakdown of ${\cal PT}$ symmetry at high energies}

Our construction of a closed form of the bound states is
transparent and, undoubtedly, potentially useful. For the straight
path ${\cal C}$ with $\omega=\varphi=0$ and for all the values of
$Z>0 $ which are not too large, the square-well model already
found interesting applications in the study of the spontaneous
${\cal PT}-$symmetry breaking at the sufficiently large $Z$
\cite{Geza}. An even more important role of this model seems to
have emerged within the supersymmetric quantum mechanics
\cite{Bagchi}. In all these and similar applications, our present
results simply mean that all the changes caused by a shift of a
small size $|\omega|$ remain smooth if an only if we do not move
to the very high energies.

In contrast to that, an introduction of {\em any} non-vanishing
shift $\omega$ changes the high-energy region completely and
abruptly. In place of infinitely many real and positive energies
$E_n(Z)$, $ n = 0, 1, \ldots$ which formed the complete spectrum
at $\omega=0$, the choice of {\em any} $\omega =\tan \varphi \neq
0$ makes the number of the real intersections $(\sigma_n,\tau_n)$
{\em finite}, $n = 0, 1, \ldots, n_{max}(Z,\varphi)$ with a
certain maximal real energy at $n_{max}(Z,\varphi)< \infty$. The
mathematical foundation of this conclusion is almost trivial: Up
to a finite number of exceptions, the energies may only be
generated by the intersections in the domain of the large $\tau
\gg 1$ where both the $Z-$dependent hyperbolas with $\omega>0$ and
$\omega = -|{\omega}| < 0$ have almost the same asymptotic
representation,
 \be
  \tau=
 -\frac{1}{|\omega|}\,\sigma \mp
 \frac{|X^2|}{(|\omega|+1/|\omega|)\sigma} +{\cal O}\left (
 \sigma^{-3}\right ), \ \ \ \ \ \ {\rm sign}\ \omega = \pm 1\,,
 \ \ \ \ \ \ \ \sigma \ll -1\,.
   \label{asksindl}
  \ee
This means that both of them share the dominant term (representing
just the straight line of their common asymptote $D_{left}$) and
approach this asymptote at an inverse-power rate from above or
below, respectively (the former case is illustrated in Figure~5
displaying just the deviation from the asymptote).

The same asymptote $D_{left}$ is further shared by both the upper
and lower envelopes $\Theta_{(\mp 1,0)}(\sigma)$ of the second,
quickly wobbling curve. Nevertheless, from definition (\ref{Om19})
we easily derive their leading-order asymptotic form
 \be
  \tau=
 -\frac{1}{|\omega|}\,\sigma
 \mp
 \frac{|\omega|+1/|\omega|}
 {\sinh \sigma}
  +{\cal O}\left (
  {\sinh^{-2} \sigma}\right ), \ \ \ \ \ \ {\rm sign}\
   \omega = \pm 1\,,
 \ \ \ \ \ \ \ \sigma \ll -1\,.
 \label{asOm19}
 \ee
This implies that the {\em quick,  exponential} decrease of {\em
both} the envelopes in eq. (\ref{asOm19}) {\em guarantees } that
the wobbling line cannot have {\em any} real intersections with
{\em neither} of the two hyperbolic $Z-$dependent curves
(\ref{asksindl}) with their {too slow}, power-law rate of approach
to the asymptote. This is illustrated in Figure 5 as a key message
of the whole construction and implies that the number of the real
energies remains finite at {\em any} non-vanishing $\omega\neq 0$
and $Z$. In the other words, infinitely many real energies which
existed at $\omega=0$ become ``lost" and ``dissolved" in complex
conjugate pairs. This occurs precisely at the moment when (say, in
Figure 5) the intersecting $Z-$dependent hyperbola moves (say, due
to a slight increase of $Z$) to the top of a particular half-oval
(there, the two energies merge at a ``Bender-Wu singularity"
\cite{BW} or ``exceptional point" \cite{Heiss}) and, in the next
stage, separates from the half-oval completely (ref. \cite{Geza}
studied this type of a pairwise complexification of the
square-well energies at $\omega = 0$ in more detail).

The disappearance of the real intersections of the two curves in
Figure 5 occurs at any $Z \neq 0$ and implies that the ${\cal PT}$
symmetry of our wavefunctions becomes broken at all the
sufficiently large energies. In the other words, our initial
choice of the form of the wavefunctions does not suddenly
represent {\em all} the possible bound-state solutions. In a way
discussed in full detail in our previous $\omega=0$ study
\cite{Geza} this means that all the ``missing" bound states must
be sought in a certain more-parametric and manifestly ${\cal
PT}-$symmetry-breaking form.

The only exception in encountered at $Z=0$ where the message
offered by Figure~5 is different because in the limit $Z \to 0$
the upper hyperbolic curve moves down and coincides {\em strictly}
with the horizontal axis. This forces us to return to the very
origin of our present construction and repeat all its steps under
the new explicit postulate that $t=0$. In this case, the kind
reader may easily verify that the $Z=0$ result {\em proves} in
fact {\em independent} of the value of $\omega$ so that all the
repeated $Z=0$ and  $\omega \neq 0$ (i.e., non-Hermitian though
still ${\cal PT}-$symmetric) construction returns us back to the
energies which {\em coincide} with the well known Hermitian
square-well spectrum.

\section{Summary \label{VI}}

All our results are summarized in Table~\ref{table1} which may be
read, first of all, as an advertisement of our almost involuntary
discovery of an extremely elementary and transparent new ${\cal
PT}-$symmetric model with real energies (cf. the last line). On a
more general level, the main item in the review Table~\ref{table1}
(viz., its last but one line) warns against all the non-critical
intuition which might prove misleading in the realm of ${\cal PT}$
symmetric models. In this sense, our results may be perceived as a
non-numerical complement to numerical experiments of paper
\cite{BBjmp} where several ``not entirely smooth" potentials
clearly inclined towards a spontaneous ${\cal PT}-$symmetry
breakdown at high energies.

On this background we believe that in the nearest future,
attention will be re-attracted to the real role of non-analyticity
in the ${\cal PT}$ symmetric potentials and models, with
inspiration by our present rigorous proof that any non-vanishing
shift of $\omega$ at $Z\neq 0$ {\em makes} the ${\cal
PT}-$symmetry of our square-well model {\em suddenly} to break
down. This breakdown involves infinitely many levels at once,
i.e., it occurs in a way which seems characteristic for virtually
all the exactly solvable {\em analytic} models
\cite{Lev,systematic}. At the same time, the discontinuity of the
breakdown might reflect its {\em non-analytic} origin, contrasting
with the robust survival of the reality of spectra under
path-deformations in many not too strongly singular analytic
potentials \cite{Jakubsky}.

We have seen that the square-well model is exceptional in
representing a solvable laboratory which seems to lie on a very
boundary between ``robust" and ``fragile" models with ${\cal
PT}-$symmetry. In this sense, our present key message is
encouraging since the geometric language of our innovated
``moving-lattice" method proved extremely efficient and seems
productive. Its former aspect becomes clear when we compare the
$\omega=0$ discussion here and in ref. \cite{SQW}, while its
second property is still to be verified in the future.

\subsection*{Acknowledgment}

Partially supported by GA AS in Prague, contract No. A 1048302.

\section*{Figure captions}

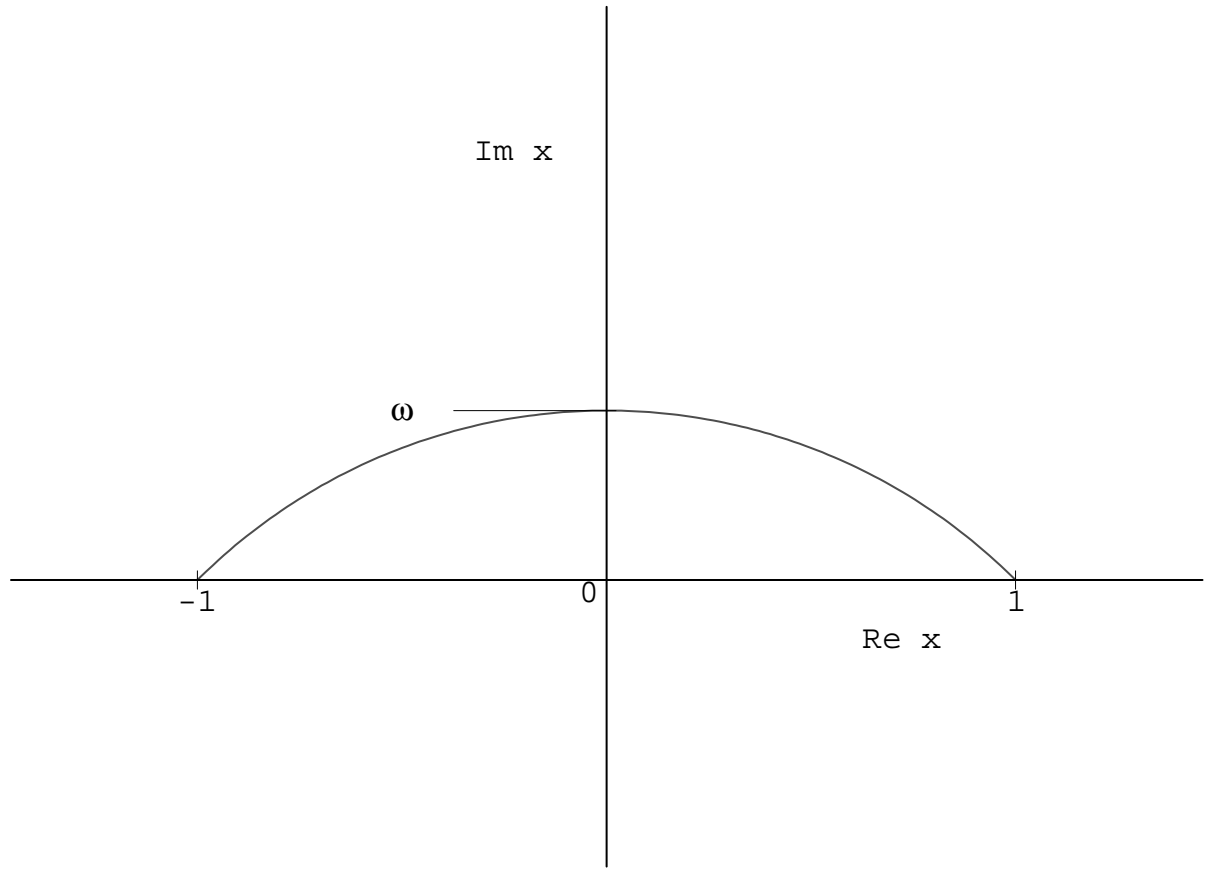
\subsubsection*{Figure 1. Integration path
${\cal C}$ with an upwards deformation
 $\omega > 0$}

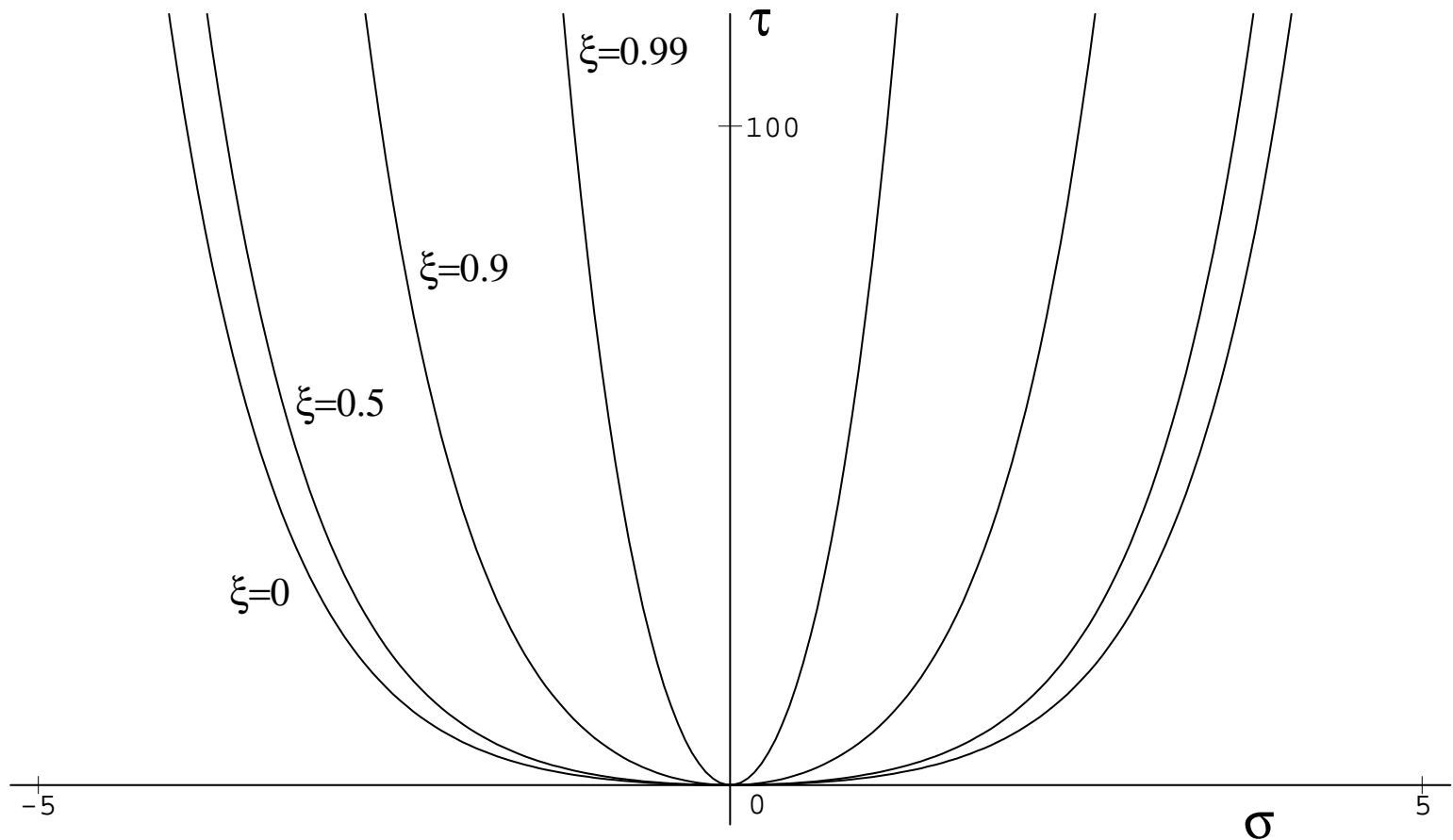
\subsubsection*{Figure 2. The lattice-dependent curve
 (\ref{quicklynu}) with $p=+1$ at a few $\xi$}

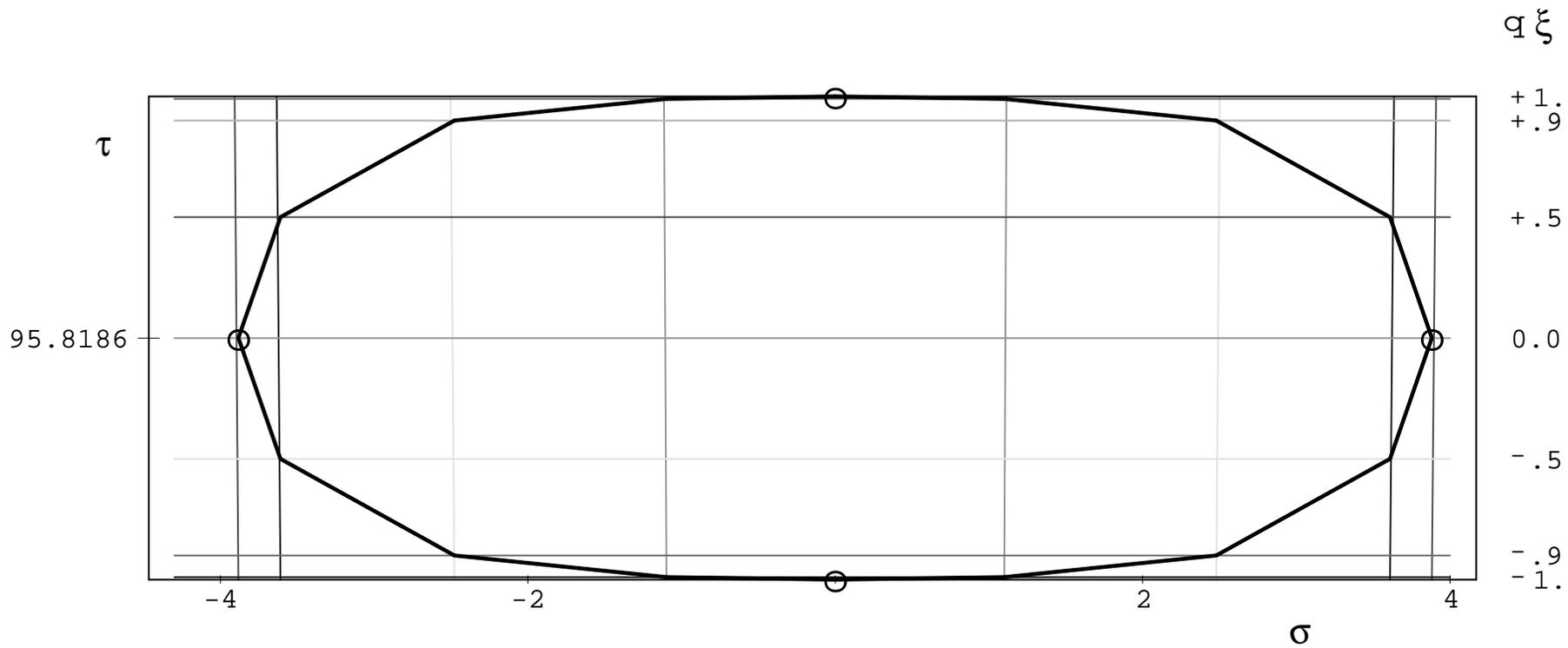
\subsubsection*{Figure 3. The  oval-shaped matching  constraint $\tau =
\Theta(\sigma)$ in its $\xi-$discreti-zation at $\omega=0$ and
$k=30$
 }

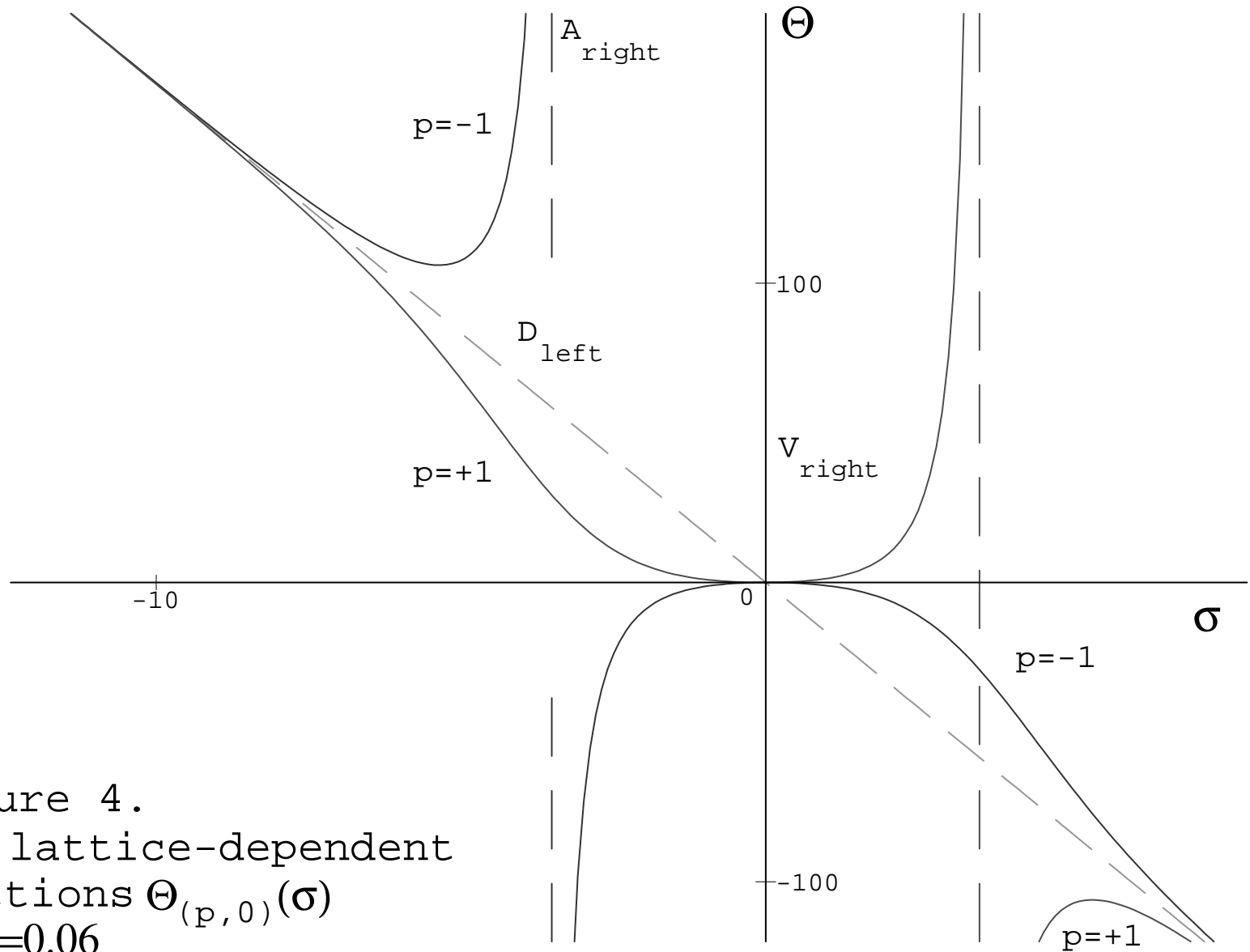
\subsubsection*{Figure 4. Two lattice-dependent functions
 $\tau=\Theta_{(p,0)}(\sigma)$ at $\omega=0.06$}

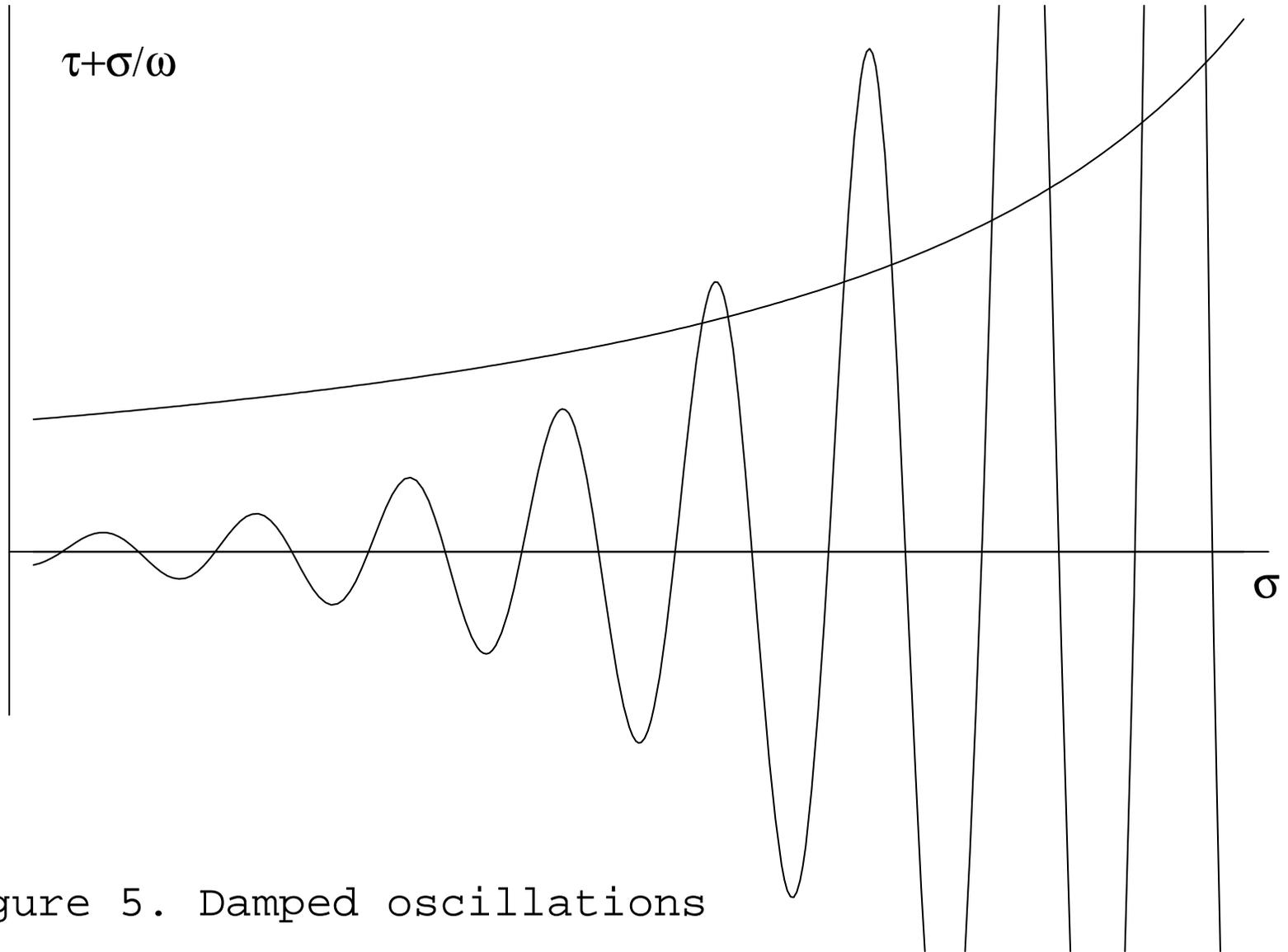
\subsubsection*{Figure 5. Damped oscillations of $\Theta(\sigma)$
vs. constraint $2st=Z$
 }

\section*{Table captions}

\subsubsection*{Table 1. Cardinalities $N^{(real)}$ and $N^{(complex)}$
 of the square-well
 energies $E$ with ${\rm Im}\,E=0$ and ${\rm Im}\,E \neq
 0$,
 respectively}

   \begin{table}

 \caption{
 Cardinalities $N^{(real)}$ and $N^{(complex)}$
 of the square-well
 energies $E$ with ${\rm Im}\,E=0$ and ${\rm Im}\,E \neq
 0$,
 respectively
} \label{table1}

  \begin{center}

  \begin{tabular}{||c|c||c|c||c||}
  \hline\hline
  $\omega$   &  $Z$ & $N^{(real)}$ & $N^{(complex)}$
 & comment\\
 \hline \hline
0&0&$\infty$ & 0 & Hermitian case \\
\hline
  0& $0 < Z <4.475...$  & $\infty$ &  0  & ${\cal PT}-$symmetric
 case of ref.
  \cite{SQW} \\
 0&  $4.475... < Z <12.8015...$  & $\infty$ &  2  &   ${\cal PT}-$symmetry
  broken \cite{Geza} \\
 0&   $Z_N < Z <Z_{N+1}$  & $\infty$ &  2N  &  ${\cal PT}-$symm.
  br.  at $E<E_{crit}(N)$ \\
  \hline
$\neq$ 0 & $\neq$ 0 & finite & $\infty$  &
${\cal PT}-$symm. br. at $E>E_{crit}(Z,\omega)$
\\
  $\neq$ 0 & 0 & $\infty$ & 0 & new ${\cal PT}-$symmetric case \\
  \hline
\hline
\end{tabular}

\end{center}

\end{table}

\end{document}